# Developments in ROOT I/O and Trees

**R. BRUN**[1], **Ph. CANAL**[2], **M. FRANK**[1], **A. KRESHUK**[1], **S. LINEV**[3], **P. RUSSO**[2], **F. RADEMAKERS**[1]

[1] CERN, Geneva, Switzerland
[2] Fermilab, Batavia, IL, United States of America
[3] GSI, Darmstadt, Germany

**Abstract**. For the last several months the main focus of development in the ROOT I/O package has been code consolidation and performance improvements. Access to remote files is affected both by bandwidth and latency. We introduced a pre-fetch mechanism to minimize the number of transactions between client and server and hence reducing the effect of latency. We will review the implementation and how well it works in different conditions (gain of an order of magnitude for remote file access). We will also briefly describe new utilities, including a faster implementation of TTree cloning (gain of an order of magnitude), a generic mechanism for object references, and a new entry list mechanism tuned both for small and large number of selections. In addition to reducing the coupling with the core module and becoming its owns library (libRIO) (as part of the general restructuring of the ROOT libraries), the I/O package has been enhanced in the area of XML and SQL support, thread safety, schema evolution, TTreeFormula, and many other areas. We will also discuss various ways, ROOT will be able to benefit from multi-core architecture to improve I/O performances.

## 1. ROOT I/O History

ROOT I/O has been as the core of ROOT since its inception and has continually improved for 12 years. This is brief review of the major enhancements. In earlier version, the I/O was requiring the user to instrument and was using streamer function either automatically generated or hand coded by the user when schema evolution was required. In 2001, we introduced support for Automatic schema evolution using the CINT dictionary information to completely drive the I/O and introduced a reference type (TRef). In 2002 and 2003 we lifted all the requirements of the user classes to be instrumented and extended all the I/O API accordingly. Since 2005 ROOT allows for files with a size larger than 2 GB. Also the compression capabilities of the I/O system for floating point umbers were improved. In 2006, we extended the TRef to work seamlessly and more efficiently with our drawing tools and improved performance over wide area network by introducing automatic prefetching.

## 2. General I/O

2.1. Major Enhancements
In addition to the enhancement described here, Leandro Franco describes in [1] the advances in and importance of the ROOT prefetching mechanism which improves by several order of magnitudes the access time to ROOT file over low latency network but also improve the access time over other network and local access. Fabrizio Furano describes in [2] the gain obtained by introducing parallel sockets connection in the I/O framework.

## 2.2. Shared Library Organization.

ROOT underwent a reorganization of its libraries in order to reduce the memory footprint of the most used functionality. In particular all the I/O classes have been gathered in their own new library, libRIO. This library includes in particular the implementation of TFile, TKey, TBufferFile, etc. The classes most used in user code, TBuffer and TDirectory are now pure abstract interfaces. TFile now derives from a new class TDirectoryFile which itself derives from TDirectory.

These changes allow the dictionary files generated by rootcint to now be totally independent of the ROOT I/O classes. If the user classes implement a custom streamer, they also need to be changed by replacing

"**Myclass::Class()->ReadBuffer(R__b,this);**"
"**Myclass::Class()->Writeuffer(R__b,this);**"

with

"**R__b.ReadClassBuffer(Myclass::Class(),this);**"
"**R__b.WriteClassBuffer(Myclass::Class(),this);**"

These changes are backward incompatible and require the user to apply the following changes:
- If you created a TDirectory object, replace it by a TDirectoryFile
- If you created a TBuffer object, replace it by a TBufferFile

## 2.3. Floating Point On Disk Compression.

We originally introduced the type "Double32_t" as a mean to use a floating point value in memory as a double and to store it on disk as a simple float, allowing the calculation to be made in double precision which using less space on disk. We extended this support to allow the user to specify, via an annotation of their class's header file, the range of value covered by their data member and, optionally, the number of significant bits the user would like to keep on disk. When these annotations are specified, the double is stored on disk as integer with the specified precision; this actually allows in some cases to keep more precision than with a float for less space, in particular due to the higher compression ratio achievable with integral values, as shown in the "Double32_t compression and precision" chart:

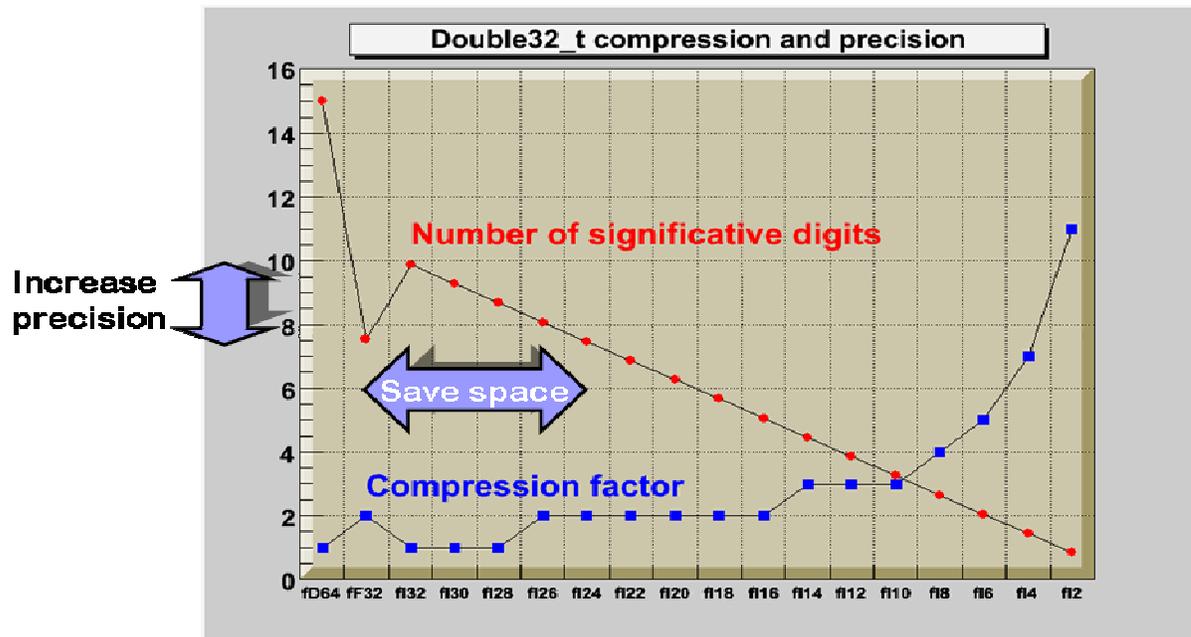

The compression factor for floating values increases with less significant digits retained.

We currently implement Double32_t as a typedef to double. Due to the way C++ handles typedef, this mean that templated type using double and Double32_t, like vector<double> or vector<Double32_t> are indistinguishable by C++ itself. This should be alleviate in the future release of the C++ standard which introduces an 'opaque typedef' that allow the target and the typedef to be different type. In the meantime, we extended the "string based" interface in ROOT to be able to distinguish those types. The "string based" interfaces include all the dictionary based I/O mechanism and the TTree branch creation mechanism that requires the user to specify the type.

We also added support for schema evolution from a container of double to the same container of Double32_t and vice et versa.

2.4. File Utilities

To support the common case where the data files are first created as a set of many files but where it is more practical to actually stored larger combined file in the long term storage areas, ROOT provides an utility named **hadd**. This utility can merge several files containing TTree instances hold the same type of data into a single file containing the concatenation of the small files. The **hadd** utility can also be used to add several histograms together. This is especially useful if the user's analysis was split over several physical machines each looking at a portion of the data and producing partial histograms in separate files.

TFile::Cp is a new static allowing the copying of any files (including non-ROOT files) via any of the many ROOT remote file access plug-ins. This simplified the implementation of new class, TFileMerger, giving access to the functionality of the utility **hadd** from within a C++ program. It allows for easy copying and merging of two or more files using the many TFile plugins (i.e. it can copy from Castor to dCache, or from xrootd to Chirp, etc.).

```
TFileMerger m;
m->AddFile("url1");
m->AddFile("url2")
m->Merge();
```

The AddFile() and Merge() use the Cp() to copy the file locally before making the merge, and if the output file is remote the merged file will be copied back to the remote destination.

TFile::Open has a new option "CACHEREAD" that first use TFile::Cp() to copy the file locally to a cache directory, then open the local cache file. If the remote file already exists in the cache, this file will be used directly, unless the remote file has changed.

A new function TFile::AsyncOpen allows for asynchronously opening a file. This call never blocks, returns an opaque handle (a TFileOpenHandle), supports a string lookup. It is currently implemented only for **xrootd** connection. This allows for a process to start opening all the needed files at the beginning of its initialization phase and while the files are opening to be able to finish the initialization. "Opening" a file is usually a relatively slow operation when the file is being retrieved from a mass storage system even if the file does not have to be retrieve from tape or any other slow medium. To alleviate this, one can now use the TFile::AysncOpen to request the opening of many file in parallel; thus avoid to have to open each of those files sequentially.

2.5. Other Consolidations

The utility **hadd** has been extended to allow the user to select the target file compression level and to select which objects (Histograms or Trees) are copied. Its performance is also enhanced using the new tree fast merge mechanism.

We introduce several thread safety tweaks, including reducing the code's reliance on gFile and gDirectory in the ROOT I/O inner code.

We also extended support for operator new in the dictionaries and implemented a proper 'destructor' for 'emulated objects'. These changes allow for proper allocation and deallocation of emulated objects in all cases.

To support the existing class structure in Geant4, we enabled I/O for C-style array of polymorphic arrays. I/O for C-style arrays of strings was also enabled. And we added support for TBuffer's operator<< and operator>> from the CINT command line.

2.6. ROOT and SQL

We extended the SQL support in ROOT by adding for ODBC and by introducing two new major facilities. SQL access directly from ROOT allows using the existing ROOT data analysis toolset even if the data is stored in a SQL database. In addition it allows correlating this same information with information stored in a ROOT file. Typically an experiment will store its evolution over time of the environment conditions measurements in an SQL database while store the event data in a ROOT file.

*2.6.1. TSQLStatement*

This new class allows the ROOT SQL interface to take advantage of the underlying RDBMS's prepared statement mechanism. It works with native data types: integer, double, date/time, string, null, etc. and introduces support for binary data (BLOBs). It is not only useful for SELECT, but also for INSERT queries. TSQLStatement is implemented for all major RDBMS: MySQL, Oracle, PostgreSQL, SapDB. It provides significant improvements in performance, especially for bulk operations, especially for Oracle (factor of 25 - 100).

*2.6.2. TFileSQL*

Allow access to table via the well known TFile interface. It supports both classes with and without custom streamer. TFileSQL will create and maintain all the tables that are needed to record the complex network of object relationship.[3]

**3. Trees**

3.1. Auto-dereferencing

TRef and TRefArray are now auto-dereferenced when used in TTree::Draw. This auto-dereferencing must be explicitly enabled at write-time of the TTree by calling TTree::BranchRef. For collections of references either a specific element of the collections may be specified or the entire collection can be scanned. The same framework can be used for any Reference classes, for example POOL::Ref.

The TTreeFormula operator @ applied to a reference object allows to access internals of the reference object itself.

The dereference mechanism also works across distributed files (if supported by the reference type)

This feature required a new abstract interface: TVirtualRefProxy. It includes a generic interface returning referenced objects and their types and support both single references and collection of references. The concrete implementation must be attached to the corresponding TClass:

**TClass::GetClass("TRef")->AdoptReferenceProxy(new TRefProxy());**

3.2. Fast Merging

We introduced a new option for the Merge and CloneTree series of function. This "fast" option request the copy operation to be done without unzipping the basket and without un-streaming the object. The raw bytes are then copied directly from one file to the other. This offers also an opportunity to re-cluster the basket within the TTree, and reduce the read time, if the read pattern is known in advance.

This new copying and merging method offers much higher performance than previous implementation that required the unzipping and un-streaming of every single data object even if the data itself was not looked at or used. However this new technique is only available for copying all the entries of a set of branches.

### 3.3. TTree Interface enhancements

A new method TTree::ReadFile, allows for the creation a TTree based on an ASCII file containing the data. The description of the file's content is described using the familiar leaflist format, for example:
   "T->ReadFile("basic.dat","x:y:z");"
The new method TTree::GetEntries returns the number of entries passing the selection.

### 3.4. TTree::Draw and TTree::Scan

TTree::Draw can now directly plot all three types of string (std::string, TString and char*).

Objects can also be easily plotted by adding a new method named either AsDouble or AsString, AsDouble being used if both are present. This is used for example to allow the plotting of the object of type TTimeStamp.

TTree::Scan was extended to allow the formatting, using printf style formatting options, of the output.

```
tree->Scan("val:flag:flag:c:cstr", "", "col=::#x:c:");

***********************************************************************
*    Row   *      val *     flag *     flag *        c *     cstr *
***********************************************************************
*        0 *        0 *        1 *      0x1 *        a *      i00 *
*        1 *      1.1 *        2 *      0x2 *        b *      i01 *
```

### 3.5. TTree::MakeProxy

The new function MakeProxy generates a skeleton analysis class inheriting from TSelector whose main features are:
- **on-demand** loading of branches
- ability to use the 'branchname' as if it was a data member
- protection against array out-of-bound
- ability to use the branch data as an **object** (when the user code is available)
- Gives access to all the functionality of TSelector

In addition, this new method has the advantage over the classical MakeSelector of generating only a header file which should not be modified by the user and includes the user generated source code. This means that unlike MakeSelector which requires the user to modify the generated files, MakeProxy can be run many times without affecting the user modifications.

Some example are located in $ROOTSYS/tutorials: *h1analysisProxy.cxx , h1analysProxy.h and h1analysisProxyCut.C*

This new technique is used to implement support for **tree->Draw("hsimple.C");** where hsimple.C is a user generate source file which contains at least a function named hsimple which can use all of the features of MakeProxy and any other arbitrary C++ constructs. The result of this function will be plotted as usual by TTee::Draw.

### 3.6. Keeping Track of List of Events

Instances of type TEvenList represents a simple sequence of the entries numbers which pass some criteria. However TEventList's implementation, an array of long long, scales linearly with the number of entries selected and is not well suited for Proof due to its monolithic nature.

The Parallel ROOT Facility, PROOF, is an extension of ROOT allowing transparent analysis of large sets of ROOT files in parallel on clusters of computers or multi-core machines. The main design goals for the PROOF system are transparency, scalability, and adaptability.

PROOF is primarily meant as an interactive alternative to batch systems for Central Analysis Facilities and departmental workgroups (Tier-2's) in particle physics experiments. However, thanks to a multi-tier architecture allowing multiple levels of masters, it can be easily adapted to a wide range of virtual clusters distributed over geographically separated domains and heterogeneous machines (GRIDs).

Apart from the pure interactive mode, PROOF has also "interactive batch" mode. With "interactive batch" the user can start very long running queries, disconnect the client and at any time, any location and from any computer reconnect to the query to monitor its progress or retrieve the results. This feature gives it a distinct advantage over purely batch based solutions that only provide an answer once all sub-jobs have been finished.

To overcome TEventList's limitations we introduced a new class, TEntryList which is scalable, modular, has a small footprint and can be loaded partially in memory. This is done by using 'block' holding information on only 64000 entries. Inside one of this block the information is stored either as a bit field or as a regular array entry number depending which is smaller. The bit field representation is the more compact representation when more than 1/16 of the entries in the block pass the selection criteria.

TEntryList objects can be stored and restored easily from files which can be independent one from the other. TEntryList objects can be combined and split to process trees independently from their chain. This feature is essential to be able to process in parallel the content of the chain without having strong ties between the processors and nodes that process each parts.

## 4. Conclusion

Even after 12 years of ROOT the I/O area is still improving. There has been were quite a number of developments including:
- Remote file performance
- SQL Support
- Tree I/O from ASCII, tree indices
- Auto dereferencing
- Fast Merge

There will certainly continue to be developments in the I/O area. However our main concern is currently to keep incrementally improving the quality and stability of the existing features. Future developments will focus on:
- Consolidation (*Thread Safety*)
- Data Model Evolution
- Increasing support for STL, including splitting collection of polymorphic objects.